# When does privatization spur entrepreneurial performance? The moderating effect of institutional quality in an emerging market

**Christopher J. Boudreaux**
Department of Economics
College of Business
Florida Atlantic University
777 Glades Road, Kaye Hall 145
Boca Raton, FL 33431 USA
cboudreaux@fau.edu

**ABSTRACT**

We explore how institutional quality moderates the effectiveness of privatization on entrepreneurs' sales performance. To do this, we blend agency theory and entrepreneurial cognition theory with insights from institutional economics to develop a model of emerging market venture performance. Using data from the World Bank's Enterprise Survey of entrepreneurs in China, our results suggest that private-owned enterprises (POEs) outperform state-owned enterprises (SOEs) but only in environments with high-quality market institutions. In environments with low-quality market institutions, SOEs outperform POEs. These findings suggest that the effectiveness of privatization on entrepreneurial performance is context-specific, which reveals more nuance than previously has been attributed.

**Abbreviations**

*SOE* = State-owned enterprise
*POE* = Private-owned enterprise

# 1. Introduction

Why are some countries more prosperous than others? This question dates back to the 18$^{th}$ century with Adam Smith's *An Inquiry into the Nature and Causes of the Wealth of Nations* (Smith, 1776). Most scholars now recognize the importance of institutions in promoting prosperity (Acemoglu, Johnson, & Robinson, 2005; North, 1990; Williamson, 2000)—especially through the channels of entrepreneurship and innovation (Acs, 2006; Acs, Audretsch, Braunerhjelm, & Carlsson, 2004; Acs & Szerb, 2007; Bjørnskov & Foss, 2016; Bradley & Klein, 2016). China presents a puzzle, however. Despite many obstacles to markets and private enterprise, entrepreneurship has flourished in China (Zhou, 2013, 2014). If a strong institutional environment influences entrepreneurship and weak institutions harm it, why has entrepreneurship in China thrived despite an environment of low-quality market institutions?

We now know that the success or failure of an emerging market depends largely on the performance of entrepreneurs (McMillan & Woodruff, 2002). More importantly, however, the success of private enterprises has varied in transition economies like Russia, China, Poland, and Vietnam. In some transition countries, the government impeded the development of private enterprises, and in other economies the government fostered an environment conducive to entrepreneurship (McMillan & Woodruff, 2002). Thus, the government can play a large role in fostering or impeding the success of these transitions (Frye & Shleifer, 1997).

Leveraging insights from new institutional economics (Acemoglu et al., 2005; Bylund & McCaffrey, 2017; North, 1990; Williamson, 2000), we propose that the effectiveness of privatization on entrepreneurial performance depends critically on the underlying quality of market institutions in an emerging market. We use agency theory (Eisenhardt, 1989; Jensen & Meckling, 1976; Ross, 1973) and entrepreneurial cognition theory (Baron, 1998; Busenitz & Lau, 1996; Wright, Hoskisson, Busenitz, & Dial, 2000) to explain how privatization is more beneficial for entrepreneurs as the quality of market institutions increases. By reducing transaction costs, privatization alters entrepreneurs' incentive structure (i.e., they become residual claimants) and consequently modifies their efforts in ways that allow entrepreneurs of private-owned enterprises (POEs) to outperform state-owned enterprises (SOEs).

Stated more formally, the purpose of our study is twofold: First, we examine how privatization affects entrepreneurs' performance in an emerging market. Second, we ask whether entrepreneurs' performance depends on the underlying institutional environment. We answer these questions by combining a sample of Chinese entrepreneurs provided by the World Bank Enterprise Survey with a measure of the quality of market institutions for Chinese provinces. Together, the dataset encompasses a wide variety of regions, industries, and provinces containing information on the quality of the institutional environment under which Chinese entrepreneurs operate. Consistent with prior research (Boudreaux, Nikolaev, & Klein, 2018; Bowen & Clercq, 2008; Estrin, Korosteleva, & Mickiewicz, 2013; McMullen, Bagby, & Palich, 2008; Stenholm, Acs, & Wuebker, 2013), our results suggest that the quality of the institutional environment matters to the extent that it moderates the effectiveness of the privatization process on entrepreneurs' sales performance. More specifically, we find that in provinces with low-quality market institutions, SOEs outperform POEs, but the outcome is reversed in provinces with high-quality market institutions. Accordingly, we speak to the ongoing conversation on the complex factors of entrepreneurial activity (Autio, Pathak, & Wennberg, 2013; Bjørnskov & Foss, 2016; Kim et al., 2016; Terjesen, Hessels, & Li, 2016) as well as the importance of context to entrepreneurship (Zahra, 2007; Zahra, Wright, & Abdelgawad, 2014)—especially in emerging markets.

Our study makes two main contributions to the literature. First, viewing the institutional context as an antecedent of entrepreneurial performance (Bowen & Clercq, 2008; De Clercq, Lim, & Oh, 2013), we draw upon agency theory (Eisenhardt, 1989; Jensen & Meckling, 1976; Ross, 1973) and entrepreneurial cognition theory (Baron, 1998; Busenitz & Lau, 1996; Wright et al., 2000) to propose that entrepreneurs' performance depends not only on the degree of privatization, but also the quality of the institutional environment. Our model suggests that the effectiveness of privatization for entrepreneur's sales depends critically on the quality of the institutional environment.

Second, we empirically examine this model using a sample of entrepreneurs in China provided by the World Bank's Enterprise Survey. This survey spans 25 regions and 24 industries encompassing a wide array of performance related questions. We combine this dataset with data from the Provincial Capital Freedom (PCF) Index developed by the China Institute of Public Affairs (CIPA) (Feng & Shoulong, 2011), which measures the quality of market institutions at the provincial-level in China. Our dataset, therefore, contains important information on the quality of the Chinese institutional environment (Ge et al., 2017), which departs from prior studies that utilize macro-level measures of market reforms (Banalieva, Eddleston, & Zellweger, 2015; Peng & Jiang, 2010). Instead, our measure of institutional quality captures the quality of the market environment across Chinese provinces. Accordingly, our model synthesizes insights from institutional economics, agency theory, and entrepreneurial cognition theory, which offers a more nuanced explanation of entrepreneurial performance in emerging markets than previous studies.

These findings have important implications. For entrepreneurs—especially those in emerging markets—our results suggests that privatizing can either help or harm entrepreneurs' sales performance, and this effect depends critically on the underlying quality of market institutions. For policy makers, the findings indicate that privatization helps entrepreneurs' performance but only in regions with strong protections of property rights, police and court protection, a stable money supply and financial sector afforded to entrepreneurs, and a small public sector relative to the private sector. If the desire of policy is to promote entrepreneurship (Acs, Åstebro, Audretsch, & Robinson, 2016; Acs & Szerb, 2007; Shane, 2008), encouraging the development of stronger underlying market institutions facilitates the effectiveness of entrepreneurs' outcomes (Boudreaux, 2014). For educators, our results shed new light on entrepreneurship in transition economies and emerging markets (McMillan & Woodruff, 2002; Park, Li, & Tse, 2006; Svejnar, 2002; Zhou, 2013) and the complex interaction between institutions and entrepreneurship (Bjørnskov & Foss, 2016; Kim et al., 2016; Terjesen et al., 2016).

## 2. Theory and hypotheses
*2.1. Entrepreneurship in emerging markets and transition economies*

Entrepreneurship plays an important role in emerging markets and transition economies like China (Ge et al., 2017; Tran, 2018; Zhou, 2013, 2014). In transition economies, the means of production were once controlled by government-directed central planners, but later transitioned toward capitalism (McMillan & Woodruff, 2002; Svejnar, 2002). Compared to planned economies dominated by large firms and few consumers goods, small and medium-sized firms emerge in transition economies (Park et al., 2006). Transition economies thus create an environment conductive to entrepreneurship where entrepreneurs supply consumer goods, mobilize savings, and compete with state-created monopoly (McMillan & Woodruff, 2002). There is heterogeneity, however, in the rate and speed of transitions both between countries and *within* countries. Regarding China, Ge et al., (2017, p. 408) explain, "Although pro-market reform policies are set by the Chinese federal government, they are implemented at various levels and speeds by the provincial governments."

China first began its transition towards privatization under the leadership of Deng Xiaoping in 1978. Reflecting on this transition (Zhou, 1996), Deng Xiaoping recalled, "All sorts of small enterprises boomed in the countryside, as if a strange army appeared suddenly from nowhere" (McMillan & Woodruff, 2002, p. 153). Prior to this transition, private enterprise was forbidden in China (Ge et al., 2017), but by 2005 China had approximately 24 million private enterprises (Loyalka, 2006) and the number of registered SME's exceeded 4.3 million by 2012 (Ministry of Commerce People's Republic of China, 2012). Despite this impressive feat, China's institutional environment still creates obstacles to private enterprise as Chinese officials continued to discourage entrepreneurship and private enterprise activity due to a reluctance to reform legal and market institutions (Zhou, 2014). For instance, the Heritage Index of Economic Freedom ranks China 110 out of 180 ranked countries and categorizes China as "mostly unfree" (Heritage, 2018). Similarly, the Economic Freedom of the World Index ranks China 112 out of 159 countries (Gwartney, Lawson, & Hall, 2017). As we will come to see, the low ranking of China's institutional environment is crucially important because the institutional environment and entrepreneurship are inexorably connected (Bruton, Ahlstrom, & Li, 2010; Pacheco, York, Dean, & Sarasvathy, 2010).

*2.2. Privatization*

Scholars often credit privatization with many successes following post-Socialist transitions (McMillan & Woodruff, 2002). Privatization propels greater enterprise restructuring and entrepreneurship activity (Djankov & Murrell, 2002), and it also allows new entrants to increase competition and erode the once substantial profits of large SOEs operationalized as monopolies (McMillan & Woodruff, 2002). However, although much attention has been given to entrepreneurs' roles in fostering economic growth and job creation (Acs, 2006; Acs & Szerb, 2007; Audretsch, Keilbach, & Lehmann, 2006; Baumol, 1986, 2002; Baumol & Strom, 2007; Bjørnskov & Foss, 2013), less attention has been given to entrepreneurs' abilities to navigate the uncertainty that emerges during economic transitions (Jackson, Klich, & Poznanska, 1999). During the Chinese post-Socialist transition, for example, entrepreneurs had to navigate a tightrope of enhancing efficiency and productivity while also satisfying state planners. But as environmental characteristics also transitioned toward a pro-market institutional environment, entrepreneurs had to adapt their strategic orientation to satisfy the market rather than state planners (Tan, 2007). We contend that POEs are naturally suited to deal with the market disequilibria (Schultz, 1975), which should be especially helpful during economic transitions within emerging markets. SOEs, however, might hesitate to invest in a riskier climate (Tan, 2001). We thus expect that POEs and SOEs respond differently during economic transitions within emerging markets.

There are several reasons to expect that entrepreneurs of POEs might outperform SOEs. First, privatization alters incentives toward productivity and efficiency. Agency theory (Eisenhardt, 1989; Jensen & Meckling, 1976; Ross, 1973) helps to explain the effect of these transforming incentives. In agency theory, managers (i.e., agents) of POEs and SOEs act in their own self-interest, rather than the self-interest of the owners (i.e., principals). Privatization helps reduce these divergent incentives by imposing external and internal control mechanisms such as markets for managers, capital, corporate control, managerial participation in ownership, reward systems, and the board of directors (Cuervo & Villalonga, 2000). The agency problem, therefore, can be resolved by optimizing the risk preferences of principals and agents, realigning their incentives, and effective monitoring (Dharwadkar, George, & Brandes, 2000). Despite their ability to realign the incentives of POEs, these mechanisms are virtually absent in SOEs. Consequently, agency theory contends that privatization can enhance performance by inducing change in corporate governance and altering managerial incentives (Bos, 1991; Cornelli & Li, 1997; Laffont & Tirole, 1993; Sappington & Stiglitz, 1987; Schmidt, 1996; Vickers & Yarrow, 1988). By transforming individuals from managers to entrepreneurs, individuals become the residual claimants of the private enterprise (Zahra et al., 2000), and residual claimants benefit from positive performance while suffering from bad performance. In contrast, managers in SOEs do not face the same profit motive (Hayek, 1945). Instead, managers of SOEs face different incentives that ultimately discourage entrepreneurial behavior (Niskanen, 1971). Empirical evidence suggests that managers of SOEs are less innovative and more risk averse than entrepreneurs of POEs in the private sector (Tan, 2001). We thus expect that entrepreneurs of POEs will increase productivity, efficiency, and profitability in an attempt to earn profits and avoid losses. Managers of SOEs do not face the same incentives.

Second, privatization transforms the scope of the organization and the responsibilities of the entrepreneur. Rooted in psychology, entrepreneurial cognition theory (Baron, 1998; Busenitz & Lau, 1996; Wright et al., 2000) suggests that individuals' cognitive heuristics influence strategic entrepreneurial decisions, which explains how entrepreneurs' expectancy beliefs persist despite receiving negative feedback (Gatewood, Shaver, Powers, & Gartner, 2002). Crucially, this resilience is important for entrepreneurship (Williams & Shepherd, 2016). Managers of SOEs, however, do not form these same resilience traits because the organization's decisions are centrally determined by the state and/or the planning board (Pelikan, 1986). In the private sector, however, entrepreneurs must make their own decisions (i.e., decentralized decisions). Entrepreneurs in POEs thus have more responsibility to determine the strategic decisions and directions the firm will take. More importantly, these cognitive heuristics are especially helpful in uncertain and rapidly changing environments as the use of heuristics allows entrepreneurs to "make sense out of uncertain and complex situations more quickly" (Wright et al., 2000, p. 593). As privatization increases, the incentives shift the focus toward efficiency and away from a managerial mindset.[1] "Privatization encourages strategies designed to shape and exploit market imperfections, garner monopolizing rents, collaborate with scarce partners, and exploit relationships with government officials" (Doh, 2000, p. 555). Entrepreneurial firms in China have adopted a strategic mindset that includes speed, stealth, and sound execution, which allows entrepreneurs to take advantage of first-mover advantages in a turbulent environment (Doh, 2000; Tan, 2001). Thus, entrepreneurs in the private sector are better suited to dealing with the uncertainties emerging through market disequilibria (Schultz, 1975). Based on these reasons, we propose the following baseline hypothesis:

**Hypothesis 1.** Privatization positively affects entrepreneurs' sales performance.

*2.3. The moderating effect of institutional quality on entrepreneurs' sales performance*

Although the institutional environment in China is often thought to be improving (Zhou, 2013, 2014), recent work shows that reforms have the opposite effect in some provinces, where the institutional environment is becoming harsher not better (Banalieva et al., 2015). Moreover, although privatization has been heralded with many successes, several exceptions highlight problems associated with emerging markets such as transitioning too quickly (McMillan & Woodruff, 2002). Privatization, thus, might encourage economic and entrepreneurship activity but its effectiveness will likely depend on the quality of the institutional environment. Because privatization has been shown to be important in transition economies (Zahra & Hansen, 2000; Zahra et al., 2000), we hypothesize that the quality of market institutions moderates the relationship between the effectiveness of privatization and entrepreneurs' sales performance.

---

[1] In centrally planned economies, for instance, decisions were often made by central planners and management only carried out routine orders on behalf of the administration (Pelikan, 1986). Despite possessing power and control (Puffer, 1994), Soviet management was inflexible and discouraged entrepreneurial behavior, which led to low value finished goods (Filatotchev, Wright, Buck, & Zhukov, 1999).

There are several reasons to expect that the quality of market institutions moderates the relationship between privatization and entrepreneurs' sales performance. In environments with low-quality market institutions, *unproductive* entrepreneurship flourishes (Baumol, 1990; Sobel, 2008). These environments incentivize entrepreneurs to substitute productive activities (e.g., innovation and price competition) for unproductive activities (e.g., rent seeking, lobbying for subsidies) and destructive activities (establishing/protecting entry barriers), which ultimately lowers productivity and economic growth (Murphy, Shleifer, & Vishny, 1991, 1993). These findings lead researchers to conclude that high-quality institutional environments nurture entrepreneurship while low-quality ones inhibit it (Banalieva et al., 2015; Bruton et al., 2010; Cuervo-Cazurra & Dau, 2009; Nikolaev, Boudreaux, & Palich, 2018). Thus, POEs are likely disadvantaged relative to SOEs in environments with low-quality market institutions.

Some entrepreneurs have thrived in China, despite the presence of low quality institutional environments (Ge et al., 2017). In these environments, political and social connections become relatively more important for entrepreneurship (Zhou, 2013, 2014). Because political connections often substitute for deficient legal and market institutions (Zhou, 2013), *who you know* becomes relatively more important than *what you know* (Boudreaux & Nikolaev, 2018). Here, a clan mentality dominates decision making (Hofstede, 2001). As a result, entrepreneurs wishing for success in such environments must build and maintain relationships with important persons in power. Empirical evidence supports this logic. Despite the weak investor protection afforded by these deficient institutions, reinvestment rates often increase for entrepreneurs who maintain good relationships with politicians (Ge et al., 2017). Due to the relationships with politicians and other important government officials, SOEs and mixed enterprises should find it easier to leverage their political and social connections. We therefore expect that entrepreneurs of POEs will underperform relative to SOEs in weak institutional contexts. In environments with high-quality market institutions, however, we expect the opposite relationship.

In high-quality institutional environments (i.e., market-enhancing), rewards are distributed through the market mechanism that rewards profits and punishes losses (Hayek, 1945). While social connections undoubtedly remain important in these environments (Boudreaux & Nikolaev, 2018), the benefits of insider connections and social ties diminishes in importance (Gwartney et al., 2017). Market-enhancing institutions promote productive entrepreneurship and discourage destructive entrepreneurship (Baumol, 1990; Boudreaux, Nikolaev, & Holcombe, 2018).

Transition economies like China highlight the relative importance of the market versus the state and the differences between POEs and SOEs. Entrepreneurs from POEs used speed, stealth, and sound executions to establish first-mover advantages when SOEs were unwilling or unable to execute similar risky decisions (Tan, 2001). Moreover, experience from Chinese SOEs reveals that they too must evolve toward a market-oriented approach if they are to survive in an institutional environment that rewards private enterprise creation (Tan, 2007).

Given these considerations, we expect that POEs and SOEs will respond differently to the changing institutional environment. In transition economies, POEs are more likely than SOEs to take risks and make investments (Tan, 2001). The market process, thus, rewards POEs in environments with high-quality market institutions. In contrast, political and social connections are relatively more important in low-quality institutional environments (Ge et al., 2017; Zhou, 2013, 2014). *Who you know* becomes more important than *what you know* in these environments (Boudreaux & Nikolaev, 2018; Hofstede, 2001). As a result, SOEs are at an advantage in environments with low-quality market institutions. Therefore, we expect that the quality of market institutions will moderate the relationship between privatization and entrepreneurs' sales performance. This leads us to the following hypothesis:

**Hypothesis 2.** POEs underperform SOEs in environments with low-quality market institutions but outperform SOEs in environments with high-quality market institutions.

## 3. Methods
*3.1. Data*

We use data from the World Bank's Enterprise Survey of Chinese entrepreneurs to test our hypotheses. This survey was conducted in 2012, spans 25 Chinese regions, 24 industries, and includes 1400 observations of which 86 percent are private enterprises, 11 percent are mixed (i.e., private and state-owned), and three percent are completely state-owned. We specifically chose this Chinese sample of entrepreneurs because we are interested in examining the composition of private and public owned enterprises. Alternatively, the World Bank provides a panel version of many different countries for its Enterprise survey, but these data cannot be compared to state-owned enterprises. The surveys were implemented following a two stage procedure. First, a screener applied a questionnaire over the phone to determine eligibility and to make appointments. Second, face-to-face interviews took place with the manager, owner, or director of each establishment.

------------------------------
Insert Table 1 about here
------------------------------

Table 1 reports the descriptive statistics. The average provincial-level of capital freedom is 7.52 (on a 10-point scale where higher numbers indicate a higher quality institutional environment). Thirty-four percent of entrepreneurs have access to a line of credit but only five percent have outstanding personal loans. Forty-three percent of all firms have funded research and development (R&D) in the past three years, 47 percent of firms have created a new product in the last year, and 59 percent of firms have created a new idea. On average, entrepreneurs work 59 hours per week and have 17 years of industry experience. Most entrepreneurs sell their product throughout China (73 percent). Only nine percent sell their product globally and 18 percent sell their product almost entirely in the local market. Lastly, it is most common for entrepreneurs to market and promote their product on a daily basis, a few times per week, or a few times per month. Only 15 percent use marketing only once in a while and 12 percent never market their product. Table 2 lists the industries covered in the study, Table 3 lists the Chinese cities covered, and Table 4 uses a non-parametric Spearman test[2] to report the correlations between the variables.

-----------------------------------------
Insert Tables 2, 3, and 4 about here
-----------------------------------------

*3.2. Measures*

*3.2.1. Entrepreneurs' sales*

Our dependent variable is entrepreneurs' sales, which we use to capture the venture's performance. Due to heteroskedasticity concerns that arise with large variances in sales, we transform our sales measure using the natural logarithm and employ standard errors that are robust to heteroskedasticity (Huber, 1967; White, 1982). The logarithmic form reduces the skewness of the variable and the transformation assigns no zeros in the measure (Wooldridge, 2015).

---

[2] We use this test, rather than the Pearson correlation, because we have a large mix of dummy and continuous variables. Correlations are similar with either method, however.

### 3.2.2. Institutional quality

To measure the quality of institutions in China, we use the Provincial Capital Freedom (PCF) Index developed by the China Institute of Public Affairs (CIPA) (Feng & Shoulong, 2011). The index consists of four areas: (1) Government and legal institutional factors, (2) Economic factors, (3) Money supply and financial development, and (4) The level of marketization in financial markets (see Table 5). Government and legal institutional factors (area 1) measure the level of government involvement in the economy. Provinces with more government involvement score lower on Area 1 of the PCF index. Economic factors (area 2) reflect the extent of private enterprise and entrepreneurial activity. Provinces with more entrepreneurial activity score higher on the PCF index. Money supply and financial development (area 3) measures macroeconomic and financial stability and liquidity in the economy. Provinces with lower rates of inflation, smaller standard deviations of inflation rates, greater bank deposits and bankcards per capita, and amounts of cash as a share of income score higher on the PCF index. Lastly, marketization of financial markets (area 4) measures the development and importance of non-state financial institutions within the banking and financial system. Provinces where financial intermediaries face greater competition and that have a greater number of non-state involvement in the stock market score higher on the PCF index. A total of 21 components aggregate together to create these four areas and the overall PCF index. The PCF index is scored on a scale from 0 to 10 where higher numbers indicate more freedom and lower numbers indicate less freedom at the provincial-level. We report more detail and the specific measurement of the components in Table 5.

-------------------------------
Insert Table 5 about here
-------------------------------

*3.2.3. Privatization*

Privatization has been shown to be an important antecedent of entrepreneurial growth (Wright et al., 2000; Zahra & Dianne Hansen, 2000; Zahra et al., 2000) and is linked to reductions in corruption (Clarke & Xu, 2004). We measure privatization as the extent that the business is owned by private interests. Specifically, we construct three category dummy variables: private enterprise (1 = 100% private ownership; 0 otherwise), state-owned (1 = 0% private ownership; 0 otherwise), and mixed ownership (1 = 1-99% private ownership; 0 otherwise). In our empirical analysis, we compare the outcomes between private ownership and the other two ownership categories.

*3.2.4. Controls*

We also include several control variables that have been shown to be related to entrepreneurial activity. We control for individual-level aspects of the financial environment that have been shown to be important antecedents of entrepreneurship (Acs & Szerb, 2007; Boudreaux & Nikolaev, 2018; De Clercq et al., 2013; Fairlie & Krashinsky, 2012; Robb & Robinson, 2014). Specifically, we include a measure for entrepreneurs' lines of credit and outstanding personal loans. The measure of credit line is dummy coded (1 = entrepreneur has access to a line of credit; 0 otherwise) and the measure of personal loans outstanding is dummy coded (1 = entrepreneur has outstanding personal loans; 0 otherwise).

There is a debate in the literature on the effect of firm size on entrepreneurship outcomes (Cooper, Woo, & Dunkelberg, 1989; Davis, Haltiwanger, & Schuh, 1998; Hall, 1986; Neumark, Wall, & Zhang, 2011). Consistent with Gibrat's law, recent contributions show that once the dynamic of firm age has been considered, the relationship between firm size and entrepreneurship is weakened (Haltiwanger, Jarmin, & Miranda, 2013). Accordingly, we control for both the size and age of the firm. Firm size is measured as the natural logarithm of the number of employees and firm age is measured as the number of years that have passed since the start of the business. In later robustness tests, we also split our sample into young and mature age groups to assess further the effects of firm age on entrepreneurs' performance.

Innovativeness is a key antecedent to business performance (Atalay, Anafarta, & Sarvan, 2013; Gunday et al., 2011; Hult, Hurley, & Knight, 2004). Research and development (R&D) has been shown to promote performance through sales (Morbey, 1988; Parasuraman & Zeren, 1983), and innovation activities such as new ideas and new products are also associated with greater performance (Artz, Norman, Hatfield, & Cardinal, 2010). We therefore include three measures of innovativeness that help capture these performance benefits. We include a measure of R&D spending that is dummy coded (1 = the firm has invested in R&D in the previous three years; 0 otherwise). We also include a measure of new product or service innovation that is dummy coded (1 = the firm introduced new products or services in the last year; 0 otherwise), and a measure of new ideas that is dummy coded (1 = the firm does research and develops ideas for new products and services; 0 otherwise).

We include several variables that account for firm-specific effort, industry-specific experience, and industry expertise that might positively influence entrepreneurs' sales performance. We include the variable, hours per week, to capture the intensity of entrepreneurial effort that has been associated with greater business performance (Fairlie & Robb, 2009). We measure this variable as the average number of hours worked on a weekly basis. It might also be important to control for entrepreneurs' industry experience. Prior experience supports new venture survival and sales (Delmar & Shane, 2006), and founders of high-growth ventures are more likely to have work experience and advanced training related to their field relative to micro-businesses (Friar & Meyer, 2003). We thus include a measure of industry experience that captures entrepreneurs' familiarity with specific industries. We measure this variable as the number of years of experience working in the industry. Franchising has also been shown to be beneficial to entrepreneurs (Combs, Michael, & Castrogiovanni, 2004). During early stages, franchisors help provide much needed resources to franchisees (Castrogiovanni, Combs, & Justis, 2006). We therefore include a measure of franchising that is dummy coded (1 = the firm is part of a franchise; 0 otherwise). Our study also includes 24 dummy indicators that encompass the different industries. Table 2 lists the industries covered in the study.

We include several measures that capture the firms' target market for their products and services. Prior research indicates that advertising can enhance profitability, especially in foreign markets (Lu & Beamish, 2004). Moreover, we expect that larger and more successful entrepreneurs will expand toward larger national and international markets, which have been shown to positively influence firm sales and growth performance (Zahra & Garvis, 2000; Zahra & Hayton, 2008). We therefore include three measures that capture firms' target market. We include a measure that captures the local market (1 = the firm sells the product mostly in the local market; 0 otherwise). We also include a measure that captures the national market (1 = the firm sells the product mostly across China; 0 otherwise). Lastly, we include a measure that captures the global market (1 = the firm sells the product mostly globally; 0 otherwise). We expect that entrepreneurs focusing their efforts in the global market will have higher sales performance than other entrepreneurs.

Product marketing has also been shown to have a beneficial impact on firm performance (Lu & Beamish, 2004; Saeed, Hwang, & Grover, 2002). Product promotion increases awareness and exposure but too much might lead to diminishing returns (Broussard, 2000). We therefore include four measures of the frequency of product marketing. We include a measure that is dummy coded if the firm never uses marketing (1 = the firm never uses marketing; 0 otherwise). We also include a measure that is dummy coded if the firm only uses marketing once in a while (1 = the firm uses marketing only once in a while; 0 otherwise). We include a measure that is dummy coded if the firm uses marketing a few times per week (1 = the firm uses marketing a few times per week; 0 otherwise). Lastly, we include a measure that is dummy coded if the firm uses marketing on a daily basis (1 = the firm uses marketing daily; 0 otherwise).

*3.3. Models*

Our models use hierarchical linear modeling to account for the multi-level structure of our data—our sample of Chinese entrepreneurs consists of 1,400 firms nested in 25 cities, nested in 12 provinces. We thus employ a multi-level regression model that combines fixed parameter estimates for our predictor variables with city-level and provincial-level random intercepts.[3] To be consistent with prior research (Ge et al., 2017), we use Variance Inflation Factor (VIF) to check for multicollinearity between independent and dependent variables. The VIFs for the independent variables ranged from 1.02 to 1.93, which is less than the "rule of thumb" of 10. This indicates no serious multicollinearity problem with the data (Greene, 2003).

We also tested for potential endogeneity bias by utilizing instruments for our variables, institutional quality and privatization. Entrepreneurs' willingness to offer and/or pay bribes (i.e., informal payments or 'gifts') is likely contingent on their perception of the quality of the institutional environment (Tonoyan, Strohmeyer, Habib, & Perlitz, 2010). If entrepreneurs operate in an environment with low-quality market institutions, they are more likely to use informal payments or gifts to help 'grease the wheels' (Bologna & Ross, 2015; Méon & Sekkat, 2005;

---
[3] We also tested the alternative random coefficients model and found that it did not significantly improve goodness of fit.

Tonoyan et al., 2010). We utilize an instrument, gifts (1 = if the firm has made any informal payments to authorities; 0 otherwise) to capture this relationship. POEs are often smaller than the larger SOEs in China (McMillan & Woodruff, 2002) and the state has more involvement in industries that relate to China's national interest (Amighini, Rabellotti, & Sanfilippo, 2013). Thus, there is likely a positive association between privatization and competition with illegal or informal enterprises. Accordingly, we utilize an instrument, informal enterprise competition (1 = if the entrepreneur responds that the enterprise competes with illegal or informal enterprises; 0 otherwise) to capture this relationship Yet, there are no clear and direct connections between these variables and entrepreneurs' sales. Following established practices in the literature (Bascle, 2008; Ge et al., 2017; Hamilton & Nickerson, 2003), we first performed the two-stage least squares regression and then used the Durbin-Wu-Hausman (DWH) test to examine endogeneity. The DWH tests for institutional quality (F = 2.01, p = 0.16; $\chi^2$ = 2.15, p = 0.14) and privatization (F = 0.168, p = 0.67; $\chi^2$ = 0.181, p = 0.68) show that the explanatory variables are exogenous. Thus, the multi-level approach is unbiased, and we report these regression results in the next section.

4. Results

We report the regression results in Table 6, which tests our hypothesis that privatization positively influences entrepreneurs' sales (Hypothesis 1), and the effectiveness of privatization on entrepreneurs' sales performance depends critically on the quality of market institutions (Hypothesis 2). Model 1 is our baseline model, which includes the direct effects of institutions and privatization and some firm-specific control variables. The results suggest that industry experience, larger target markets, larger and older firms, and entrepreneurial effort all contribute to greater sales to entrepreneurs. We find no evidence, however, that the quality of market institutions or privatization affect entrepreneurs' sales performance. We thus find no evidence to support hypothesis 1. Model 2 augments this baseline to include the interaction term, which tests our moderating hypothesis (Hypothesis 2). The findings indicate that privatization is associated with lower sales ($\beta = -1.501$; $p < 0.10$), but this effect decreases as the quality of market institutions increases ($\beta = 0.18$; $p < 0.10$). In other words, POEs have worse sales than SOEs in provinces with low-quality market institutions but greater sales in provinces with high-quality market institutions. Thus, we find support for hypothesis 2.

------------------------------
Insert Table 6 about here
------------------------------

Although we have found support for our second hypothesis, it is possible that our results are confounded by other important activities that affect privatization or sales performance. We thus account for this possibility by including several additional control variables that might plausibly influence our results. Models 3-6 of Table 5 augment our previous model to include additional controls related to innovation (Model 4), financing (Model 5), and marketing, sales, and market orientation (Model 6). We observe that innovation activities— introducing new products, new ideas, and investing in R&D—are associated with greater sales performance. We also observe that some sources of financing, such as possessing a line of credit, are associated with greater sales performance whereas possessing personal debt has a negative but insignificant effect on sales performance. Lastly, we find that franchises and greater marketing intensity are associated with greater sales performance and competition has a negative but insignificant effect on sales performance. Perhaps more importantly, the moderating effect remains and is robust to the inclusion of these additional controls. The robustness of this relationship increases our confidence that our moderating effect is not driven by other explanations for entrepreneurs' sales or privatization.

4.1. Stratified results by firm age

We also examine whether the age of the firm affects our results. While several studies have found that small firms are associated with high-growth entrepreneurship and job creation (Birch, 1979; Birley, 1986, 1987; Neumark et al., 2011), recent evidence suggests that the age of the firm actually drives this result (Haltiwanger et al., 2013). In other words, young high-growth ventures—rather than small businesses—drive employment and job creation. As a result, it is the age of the firm that matters more than the size of the firm. While we believe it is important to study a variety of different types of entrepreneurs including *everyday entrepreneurs* (Welter, Baker, Audretsch, & Gartner, 2017), we also recognize that young start-ups are often the focus of entrepreneur research. To account for this, we split our sample into two age categories: young ($\leq$ 5 years) and mature ($\geq$ 10 years). We then compare the results of each subsample to the entire sample.

-------------------------------
Insert Table 7 about here
-------------------------------

Overall, while we uncover a similar pattern for both age groups, the evidence is strongest for young firms ($\leq$ 5 years of age) since the moderating effect is strongly statistically significant ($\beta = 0.96$; $p < 0.001$). Although the moderating effect is also statistically significant for the mature firm age group ($\geq$ 10 years), the evidence is only marginally significant at the 10 percent level ($\beta = 0.19$; $p < 0.10$). While statistical significance is important, it is perhaps more important to discuss the magnitude of these differences. We can accomplish this task by discussing the moderating effects at one standard deviation above and below the mean level of capital freedom—our measure of institutional quality. For instance, in provinces with capital freedom one standard deviation above the mean level, young POEs have 74 percent higher sales than SOEs and mixed enterprises[4], but the sales of mature POEs are only 30 percent higher than SOEs and mixed enterprises.[5] However, in provinces with capital freedom one standard deviation below the mean-level, young POEs have 140 percent lower sales compared to SOEs and mixed enterprises but mature POEs only have 13 percent lower sales than SOEs and mixed enterprises. Our evidence thus suggests that private enterprise entrepreneurs have greater sales performance but only in environments with high-quality market institutions. Moreover, in environments with low-quality market institutions, private enterprise entrepreneurs have much lower sales compared to SOEs and mixed enterprises. This is especially true with respect to the young age group that is most consistent with entrepreneurship.

To better understand and interpret these results, we plot the moderating effect in Figure 1. We observe there is a negative relationship between capital freedom (i.e., institutional quality) and entrepreneurs' sales for SOEs. In contrast, there is a positive relationship between institutional quality and entrepreneurs' sales for POEs. That is, SOEs have greater sales performance in provinces with low-quality market institutions but worse sales performance in provinces with high-quality market institutions. This finding reinforces our support for our second hypothesis.

-------------------------------
Insert Figure 1 about here
-------------------------------

We also uncover other differences between the two age groups that we believe are worthy of additional discussion. Young firms have higher sales when they sell their product primarily to the local market and have worse sales when they sell products globally. Mature firms, on the other hand, have higher sales as the size of the market increases. Moreover, marketing intensity appears to help mature firms but hurt (or at least not help) young firms. Here, we observe that only marketing products every once in a while corresponds with greater sales performance for young firms, but marketing products on a daily basis corresponds with greater sales performance for mature firms. As a result, we believe this evidence points toward a liability of newness effect on marketing and promotion activities for early-stage ventures (Stinchcombe, 1965). That is, expanding the size of the market too quickly and promoting the product too intently can be detrimental to new venture performance and ultimately growth. We also find that different aged ventures experience different effects on venture performance depending on the source of financing. Young firms with more personal loans have higher sales whereas mature firms with larger credit lines have higher sales. This might reflect that credit lines take time to establish and repay the debt. Lastly, we also observe that franchising helps mature firms to increase sales but has no effect on young firms.

---

[4] This result is found by multiplying the moderating effect (0.96) by the level of capital freedom that is one standard deviation above the mean level (Mean = 7.52; SD = 1.12). That is, 0.96 x (7.52 + 1.12) = 8.29. We then subtract this amount from the coefficient on the variable, *private enterprise*: -7.556 + 8.29 = 0.738 or 74%.

[5] We perform the same calculation as the previous footnote but use the corresponding values in column 3 of Table 7.

## 5. Discussion and conclusion

By leveraging insights from new institutional economics, agency theory, and entrepreneurial cognition theory, we propose that the quality of market institutions moderates the relationship between privatization and entrepreneurs' sales performance. We combine the World Bank's Enterprise Survey data for Chinese entrepreneurs with data from the Provincial Capital Freedom (PCF) Index developed by the China Institute of Public Affairs (CIPA) (Feng & Shoulong, 2011) to show that the effectiveness of privatization on entrepreneurs' sales performance depends critically on the underlying quality of market institutions. Specifically, we find that POEs underperform SOEs in environments with low-quality market institutions but outperform SOEs in environments with high-quality market institutions. These findings suggest that the success of privatization depends ultimately on the quality of the underlying institutional environment.

These findings have important implications for both theoretical and practical reasons. First, our results are important for entrepreneurs. Although privatization can alter organizational cultures, promote risk taking, and spur innovation and entrepreneurship (Tan, 2001; Zahra & Hansen, 2000), our results suggest that private enterprise activity will be more successful in environments with market-enhancing institutions. In the absence of these formal institutions, political and social connections become relatively more important. Our study thus highlights the importance of placing the evidence in the appropriate context (Zahra, 2007; Zahra et al., 2014). Entrepreneurs can benefit from this study by learning about the importance of the institutional context. This is especially true for those who are considering foreign direct investment in emerging markets and who might be unfamiliar with these environments.

Second, our findings are important for policy makers. Because low-quality institutional environments encourage unproductive entrepreneurship and discourage productive entrepreneurship (Baumol, 1990; Sobel, 2008), policy makers can advocate for market reforms which might improve the quality of market institutions and ultimately lead to greater productivity and growth (Baumol, 1986, 2002; Baumol & Strom, 2007). Caution is needed, however, because research identifies that gradual changes are more successful than rapid privatization efforts (Spicer, McDermott, & Kogut, 2000). Nevertheless, if the desire of policy is to promote entrepreneurship (Acs et al., 2016; Acs & Szerb, 2007; Shane, 2008), privatization can be a viable strategy to achieve these goals—if the underlying environment is supportive of entrepreneurship and private enterprise activity.

Third, our results have important implications for educators. The primary contribution of our study shows that the relative success of privatization depends critically on the underlying quality of market institutions. These findings shed new light on entrepreneurship in transition economies (McMillan & Woodruff, 2002; Park, Li, & Tse, 2006; Svejnar, 2002; Zhou, 2013). More importantly, however, we add to the literature that argues for a more nuanced understanding of the relationship between institutions and entrepreneurship (Audretsch, Belitski, & Desai, 2018; Boudreaux, Nikolaev, & Klein, 2018; Estrin et al., 2013; Stenholm et al., 2013). The interaction between institutions and entrepreneurship is complex (Bjørnskov & Foss, 2016; Kim et al., 2016; Terjesen et al., 2016), and our findings, thus, are consistent with recent advances in entrepreneurship that offered nuanced approaches in transition economies and emerging markets (Ge et al., 2017; Tran, 2018).

*5.1. Limitations and future research*

As any empirical study, we face a number of limitations. We chose to examine a sample of Chinese entrepreneurs in an attempt to increase our understanding of the complex interaction between institutional environments, entrepreneurship, and privatization in a transition economies setting. While consistent with prior research in transition and emerging markets (Ge et al., 2017; Puffer, McCarthy, & Boisot, 2010), our findings apply only to entrepreneurs in China and caution should be given to the generalizability of our results. Future research, thus, should examine these relationships in alternative transition economies and emerging markets to determine the external validity of our findings. With that said, recent advances highlight that institutions have profound effects in other transition economies like Vietnam (Tran, 2018), and although this study examines the relative institutional dynamics on entrepreneurship for both new entrants and incumbents, we find these results encouraging for the applicability of institutions and entrepreneurship in alternative settings.

Relatedly, while our sample takes advantage of important regional and industrial heterogeneities in entrepreneurship, it only includes a single-year of observation. To alleviate some of the associated empirical problems with cross-sectional data, we employed a multi-level hierarchical linear model that incorporates city-level and province-level random effects. This methodology allows us to compare the performances of entrepreneurs to *similar* entrepreneurs *within* the same industry and geographic context (Audretsch et al., 2018). We therefore feel reasonably assured that our results are not driven by the omission of key confounders or otherwise important sources of regional heterogeneity. Future research, however, could improve upon our study by incorporating longitudinal designs that not only include important measures of regional and industrial classification but also incorporate multiple observations for the *same* organization over time. This research design would offer researchers an opportunity to enhance our understanding of the intricacies involved in the privatization process. For instance, research indicates that the *speed* of institutional reform (Banalieva et al., 2015) and the *speed* of privatization is an important consideration for the success of entrepreneurs in transitioning settings (Spicer et al., 2000). Future research could extend this logic to address whether slow and gradual improvements in the privatization process offer more benefits than a rapidly changing privatization that creates uncertainty during reform.

In sum, we find that POEs underperform SOEs in environments with low-quality market institutions but outperform SOEs in environments with high-quality market institutions. Our results thus suggest that the quality of market institutions moderates the relationship between privatization and entrepreneurs' sales performance. These findings highlight the importance of context to entrepreneurial performance, especially in transition economies and emerging markets.

**Table 1**

Measures and descriptive statistics.

| Variables | Measures | Mean | SD |
|---|---|---|---|
| Entrepreneurs' sales | Natural logarithm of sales | 16.88 | 1.62 |
| Private-owned enterprise (POE) | 1= if the firm is privately owned; 0 otherwise | 0.86 | 0.35 |
| State-owned enterprise (SOE) | 1= if the firm is state owned; 0 otherwise | 0.03 | 0.18 |
| Mixed ownership | 1= if the firm is partly owned by private and state interests; 0 otherwise | 0.11 | 0.31 |
| Capital freedom | Provincial Capital Freedom (PCF) index developed by the China Institute of Public Affairs (CIPA). The index consists of four areas: (1) Government and legal institutional factors, (2) Economic factors, (3) Money supply and financial development, and (4) The level of marketization in financial markets | 7.52 | 1.12 |
| Financing | | | |
|   Credit line | 1= if entrepreneur has access to a line of credit; 0 otherwise | 0.34 | 0.47 |
|   Personal loans outstanding | 1= if entrepreneur has any outstanding personal loans; 0 otherwise | 0.05 | 0.21 |
| Firm size | Natural logarithm of the number of employees | 3.36 | 1.23 |
| Firm age | The number of years that have passed since the start of the business | 12.94 | 7.84 |
| Innovative activities | | | |
|   R&D spending | 1= if the firm has invested in R&D in the past 3 years; 0 otherwise | 0.43 | 0.49 |
|   New product | 1= if the firm has sold a new product in the last year; 0 otherwise | 0.47 | 0.50 |
|   New idea | 1= if the firm has thought of a new idea in the last year; 0 otherwise | 0.59 | 0.49 |
| Hours per week | Number of hours worked per week | 58.73 | 24.3 |
| Industry experience | Number of years of experience working in the industry | 17.08 | 7.50 |
| Competition | On a scale from 0 to 4, respondents rate whether competition is an obstacle to daily operations: 0 = no obstacle; 1 = minor obstacle; 2 = moderate obstacle; 3 = severe obstacle; 4 = very severe obstacle | 0.84 | 0.87 |
| Franchise | 1= if the firm is considered an establishment that is part of a larger organization; 0 otherwise | 0.11 | 0.31 |
| Target market | | | |
|   Product is sold mostly in local market | 1= if the firm is sold mostly in the local market; 0 otherwise | 0.18 | 0.38 |
|   Product is sold mostly in national market | 1= if the firm is sold mostly in China; 0 otherwise | 0.73 | 0.44 |
|   Product is sold mostly globally | 1= if the firm is sold mostly in global markets; 0 otherwise | 0.09 | 0.29 |
| Frequency of product marketing | | | |
|   Never | 1= if the firm never uses marketing; 0 otherwise | 0.12 | 0.33 |
|   Once in a while | 1= if the firm uses marketing only once in a while; 0 otherwise | 0.15 | 0.36 |
|   A few times per month | 1= if the firm uses marketing a few times per month; 0 otherwise | 0.19 | 0.40 |
|   A few times per week | 1= if the firm uses marketing a few times per week; 0 otherwise | 0.24 | 0.43 |
|   Daily | 1= if the firm uses marketing daily; 0 otherwise | 0.30 | 0.46 |
| Industries | Twenty-three dummy variables capturing twenty-four industries | | |

*Notes.* Industries listed in Table 2.

**Table 2**
List of industries included in the study

| Industries |
|---|
| 1. Textiles |
| 2. Garments |
| 3. Leather |
| 4. Wood |
| 5. Paper |
| 6. Recorded media |
| 7. Refined petroleum product |
| 8. Chemicals |
| 9. Plastics & rubber |
| 10. Non-metallic mineral products |
| 11. Basic metals |
| 12. Fabricated metal products |
| 13. Machinery and equipment |
| 14. Electronics |
| 15. Precision instruments |
| 16. Transport machines |
| 17. Furniture |
| 18. Recycling |
| 19. Construction |
| 20. Motor vehicle services |
| 21. Wholesale |
| 22. Retail |
| 23. Information technology |
| 24. Transport services |

**Table 3**
List of cities included in the study

| Cities |
| --- |
| 1. Hefei City |
| 2. Beijing |
| 3. Guangzhou City |
| 4. Shenzhen City |
| 5. Foshan City |
| 6. Dongguan City |
| 7. Shijiazhuang City |
| 8. Tangshan City |
| 9. Zhengzhou City |
| 10. Luoyang City |
| 11. Wuhan City |
| 12. Nanjing City |
| 13. Wuxi City |
| 14. Suzhou City |
| 15. Nantong City |
| 16. Shenyang City |
| 17. Dalian City |
| 18. Jinan City |
| 19. Qingdao City |
| 20. Yantai City |
| 21. Shanghai |
| 22. Chengdu City |
| 23. Hangzhou City |
| 24. Ningbo City |
| 25. Wenzhou City |

**Table 4**

Correlation matrix of variables

| Variables | | [1] | [2] | [3] | [4] | [5] | [6] | [7] | [8] | [9] | [10] | [11] | [12] | [13] | [14] | [15] | [16] | [17] | [18] | [19] | [20] | [21] | [22] | [23] | [24] |
|---|---|---|---|---|---|---|---|---|---|---|---|---|---|---|---|---|---|---|---|---|---|---|---|---|---|
| Entrepreneurs' sales | [1] | 1 | | | | | | | | | | | | | | | | | | | | | | | |
| Capital freedom | [2] | 0.05 | 1 | | | | | | | | | | | | | | | | | | | | | | |
| Private owned enterprise | [3] | -0.10* | 0.00 | 1 | | | | | | | | | | | | | | | | | | | | | |
| State owned enterprise | [4] | 0.09* | 0.03 | -0.47* | 1 | | | | | | | | | | | | | | | | | | | | |
| Mixed ownership | [5] | 0.06* | -0.03 | -0.85* | -0.07* | 1 | | | | | | | | | | | | | | | | | | | |
| Industry experience | [6] | 0.19* | 0.02 | 0.08* | -0.02 | -0.08* | 1 | | | | | | | | | | | | | | | | | | |
| Target market | | | | | | | | | | | | | | | | | | | | | | | | | |
| Product sold mostly locally | [7] | -0.28* | 0.06* | 0.08* | -0.03 | -0.08* | -0.08* | 1 | | | | | | | | | | | | | | | | | |
| Product sold mostly nationally | [8] | 0.18* | -0.09* | -0.03 | -0.06* | 0.07* | 0.04 | -0.77* | 1 | | | | | | | | | | | | | | | | |
| Product sold mostly globally | [9] | 0.09* | 0.07* | -0.06* | 0.13* | -0.01 | 0.05 | -0.15* | -0.52* | 1 | | | | | | | | | | | | | | | |
| Franchise | [10] | 0.25* | -0.02 | -0.12* | 0.12* | 0.06* | 0.05 | -0.05* | 0.01 | 0.05 | 1 | | | | | | | | | | | | | | |
| Competition | [11] | -0.07* | 0.12* | 0.14* | -0.02 | -0.15* | 0.03 | 0.16* | -0.10* | -0.06* | -0.03 | 1 | | | | | | | | | | | | | |
| Frequency of marketing | | | | | | | | | | | | | | | | | | | | | | | | | |
| Never market product | [12] | -0.15* | -0.02 | -0.17* | -0.03 | 0.21* | -0.18* | 0.04 | -0.02 | -0.01 | -0.08* | -0.19* | 1 | | | | | | | | | | | | |
| Once in a while | [13] | -0.07* | -0.07* | 0.03 | -0.02 | -0.02 | 0.04 | 0.06* | -0.06* | 0.01 | 0.00 | 0.03 | -0.16* | 1 | | | | | | | | | | | |
| A few times per month | [14] | -0.08* | -0.10* | 0.09* | -0.03 | -0.09* | 0.01 | 0.05 | 0.01 | -0.09* | -0.01 | 0.09* | -0.18* | -0.21* | 1 | | | | | | | | | | |
| A few times per week | [15] | 0.04 | -0.01 | 0.03 | 0.05 | -0.06* | 0.03 | -0.07* | 0.02 | 0.07* | 0.03 | 0.02 | -0.21* | -0.23* | -0.27* | 1 | | | | | | | | | |
| Daily | [16] | 0.19* | 0.17* | -0.01 | 0.02 | -0.01 | 0.06* | -0.05 | 0.04 | 0.01 | 0.04 | 0.02 | -0.24* | -0.27* | -0.32* | -0.36* | 1 | | | | | | | | |
| Innovative activities | | | | | | | | | | | | | | | | | | | | | | | | | |
| New product | [17] | 0.19* | -0.07* | 0.00 | 0.00 | 0.00 | 0.06* | -0.04 | 0.02 | 0.03 | 0.12* | 0.14* | -0.29* | -0.13* | -0.02 | 0.11* | 0.22* | 1 | | | | | | | |
| New ideas | [18] | 0.22* | 0.06* | -0.11* | 0.00 | 0.12* | 0.02 | -0.13* | 0.09* | 0.05 | 0.15* | -0.05 | -0.05 | -0.19* | -0.08* | 0.06* | 0.20* | 0.29* | 1 | | | | | | |
| R&D spending | [19] | 0.32* | 0.05 | -0.03 | 0.01 | 0.03 | 0.14* | -0.10* | 0.06* | 0.04 | 0.13* | 0.07* | -0.22* | -0.14* | -0.05 | 0.05 | 0.27* | 0.51* | 0.29* | 1 | | | | | |
| Financing | | | | | | | | | | | | | | | | | | | | | | | | | |
| Credit line | [20] | 0.30* | 0.26* | -0.03 | 0.04 | 0.01 | 0.12* | -0.03 | -0.03 | 0.08* | 0.09* | 0.03 | -0.13* | -0.02 | -0.01 | 0.00 | 0.12* | 0.13* | 0.17* | 0.22* | 1 | | | | |
| Personal debt outstanding | [21] | -0.03 | 0.03 | 0.06* | -0.04 | -0.04 | -0.02 | 0.04 | -0.02 | -0.01 | -0.01 | 0.06* | 0.02 | 0.03 | -0.03 | -0.02 | 0.01 | 0.01 | 0.03 | -0.01 | 0.09* | 1 | | | |
| Firm size | [22] | 0.54* | 0.03 | -0.15* | 0.08* | 0.13* | 0.15* | -0.19* | 0.08* | 0.14* | 0.19* | -0.06* | -0.08* | -0.01 | -0.08* | 0.00 | 0.14* | 0.09* | 0.18* | 0.20* | 0.26* | -0.03 | 1 | | |
| Firm age | [23] | 0.13* | 0.02 | -0.01 | -0.06* | 0.05 | 0.39* | -0.04 | 0.07* | -0.06* | -0.02 | 0.02 | -0.03 | -0.02 | -0.02 | -0.03 | 0.08* | 0.04 | 0.06* | 0.06* | 0.06* | -0.03 | 0.02 | 1 | |
| Hours per week | [24] | 0.06* | -0.27* | 0.08* | -0.03* | -0.08* | -0.03 | -0.04 | 0.07* | -0.06* | -0.07* | 0.02 | -0.08* | -0.03 | 0.04 | 0.03 | 0.02 | 0.16* | 0.03 | 0.09* | -0.04 | 0.02 | 0.00 | -0.05* | 1 |

*Note* – Spearman non-parametric correlation matrix used. N = 1400 observations. $^*p < 0.05$.

**Table 5**
Components and Areas of the Provincial Capital Freedom (PCF) index

| Areas | | Measures |
|---|---|---|
| **Area 1** | **1** | **Government and Institutional Factors** |
| Component | 1.a | Share of Market Allocation of Resources (% of Government Expenditure to GDP) |
| Component | 1.b | % of Government Subsidies to Enterprises to GDP |
| Component | 1.c | Non-tax Burden of Enterprises |
| Component | 1.d | Local Protectionism |
| Component | 1.e | Legal Protection |
| **Area 2** | **2** | **Economic Factors** |
| Component | 2.a | Number of Enterprises Per Capita |
| Component | 2.b | Size of Non-State Sector |
| Sub-Component | 2.b.1 | Share of the Non-State Industry in Total Production Value of the Total Industry |
| Sub-Component | 2.b.2 | Share of the Non-State Sector in Total Capital Construction Investment |
| **Area 3** | **3** | **Money Supply and Financial Development** |
| Component | 3.a | Total Deposits as a Percentage of GDP |
| Component | 3.b | Inflation Rate |
| Component | 3.c | Standard Deviation in Inflation Rate |
| Component | 3.d | Share of Return from Asset of Urban Households in Their Total Disposable Income |
| Component | 3.e | Share of Cash Held by Urban Households in Their Total Disposable Income |
| Component | 3.f | Number of Bank Cards Per Capita |
| **Area 4** | **4** | **Level of Marketization of Financial Markets** |
| Component | 4.a | Competition Among Financial Intermediaries |
| Sub-Component | 4.a.1 | Percentage of Deposits with Non-State Financial Institutions to Total Deposits |
| Sub-Component | 4.a.2 | Percentage of Loans for Non-State Enterprises to Total Loans Granted by Financial Institutions |
| Component | 4.b | Stock Market |
| Sub-Component | 4.b.1 | Share of Number of Non-State Controlled Listed Companies in Total Number of Listed Companies |
| Sub-Component | 4.b.2 | Share of Number of Tradable Stocks in Total Number of Stocks |
| Sub-Component | 4.b.3 | Share of Non-State Controlled Listed Companies in Total Assets of All Listed Companies |
| Sub-Component | 4.b.4 | Share of Non-state Controlled Listed Companies in Total Funds Raised in Stock Market by All Listed Companies |

**Table 6.** Results of regression models on entrepreneurs' sales

|  | Entrepreneurs' sales | | | | |
|---|---|---|---|---|---|
|  | (1) | (2) | (3) | (4) | (5) |
| PCF index | 0.117 | -0.044 | -0.055 | -0.143 | -0.160 |
|  | (0.07) | (0.12) | (0.12) | (0.12) | (0.12) |
| Private enterprise | -0.151 | -1.501* | -1.368* | -1.623** | -1.731** |
|  | (0.10) | (0.78) | (0.76) | (0.75) | (0.74) |
| Private enterprise × PCF index |  | 0.180* | 0.163* | 0.198** | 0.218** |
|  |  | (0.10) | (0.10) | (0.10) | (0.10) |
| Industry experience (years) | 0.023*** | 0.023*** | 0.018*** | 0.017*** | 0.016*** |
|  | (0.01) | (0.01) | (0.01) | (0.00) | (0.00) |
| Target Market? (Reference category = local market) | | | | | |
| Main product sold mostly across China | 0.535*** | 0.535*** | 0.461*** | 0.453*** | 0.447*** |
|  | (0.09) | (0.09) | (0.09) | (0.09) | (0.09) |
| Main product sold mostly globally | 0.569*** | 0.566*** | 0.430*** | 0.415*** | 0.399*** |
|  | (0.15) | (0.15) | (0.14) | (0.14) | (0.14) |
| Firm size | 0.719*** | 0.720*** | 0.680*** | 0.652*** | 0.617*** |
|  | (0.03) | (0.03) | (0.03) | (0.03) | (0.03) |
| Firm age | 0.014*** | 0.014*** | 0.016*** | 0.016*** | 0.015*** |
|  | (0.00) | (0.00) | (0.00) | (0.00) | (0.00) |
| Hours per week | 0.005** | 0.005** | 0.003 | 0.003 | 0.003* |
|  | (0.00) | (0.00) | (0.00) | (0.00) | (0.00) |
| Innovative activities | | | | | |
| New product |  |  | 0.167** | 0.138* | 0.064 |
|  |  |  | (0.08) | (0.08) | (0.08) |
| New ideas |  |  | 0.168** | 0.150** | 0.094 |
|  |  |  | (0.08) | (0.08) | (0.08) |
| R&D spending in last 3 years |  |  | 0.485*** | 0.429*** | 0.389*** |
|  |  |  | (0.08) | (0.08) | (0.08) |
| Sources of financing | | | | | |
| Credit line |  |  |  | 0.497*** | 0.484*** |
|  |  |  |  | (0.08) | (0.08) |
| Personal loans outstanding |  |  |  | -0.090 | -0.044 |
|  |  |  |  | (0.15) | (0.15) |
| Competition is an obstacle? |  |  |  |  | -0.028 |
|  |  |  |  |  | (0.04) |
| Marketing and sales orientation | | | | | |
| Franchise |  |  |  |  | 0.651*** |
|  |  |  |  |  | (0.11) |
| Marketing intensity (reference category = never) | | | | | |
| Rarely (once in a while) |  |  |  |  | 0.186 |
|  |  |  |  |  | (0.13) |
| Sometimes (few times per month) |  |  |  |  | 0.251* |
|  |  |  |  |  | (0.13) |
| Frequently (few times per week) |  |  |  |  | 0.340*** |
|  |  |  |  |  | (0.13) |
| All the time (daily) |  |  |  |  | 0.379*** |
|  |  |  |  |  | (0.13) |
| Log-likelihood | -2291.03 | -2289.50 | -2251.61 | -2232.03 | -2208.50 |
| LR test vs baseline model | --- | *** | *** | *** | *** |
| Wald chi$^2$ of model fit | 980*** | 985*** | 1119*** | 1193*** | 1282*** |
| LR test vs single-level model | 53*** | 52*** | 58*** | 73*** | 68*** |

*Note* – Dependent variable is log of sales. Models estimated using hierarchical linear modeling. Standard errors in parentheses. N = 1400 observations. Industry fixed effects included in all models. * p<0.10, ** p<0.05, *** p<0.01

**Table 7.** Regression results stratified by firm age

| | Entrepreneurs' sales | | | | | |
|---|---|---|---|---|---|---|
| | Full sample | | Firms ≤ 5 years of age | | Firms ≥10 years of age | |
| | (1) | | (2) | | (3) | |
| PCF index | -0.160 | (0.12) | -0.839*** | (0.31) | -0.168 | (0.13) |
| Private enterprise | -1.731** | (0.74) | -7.556*** | (2.25) | -1.376 | (0.89) |
| Private enterprise × PCF index | 0.218** | (0.10) | 0.960*** | (0.31) | 0.194* | (0.12) |
| Industry experience (years) | 0.016*** | (0.00) | 0.032* | (0.02) | 0.007 | (0.01) |
| Target Market? (Reference category = local market) | | | | | | |
|   Main product sold mostly across China | 0.447*** | (0.09) | -0.394 | (0.32) | 0.589*** | (0.11) |
|   Main product sold mostly globally | 0.399*** | (0.14) | -0.835* | (0.47) | 0.625*** | (0.18) |
| Firm size | 0.617*** | (0.03) | 0.984*** | (0.09) | 0.562*** | (0.04) |
| Firm age | 0.015*** | (0.00) | -0.040 | (0.10) | 0.018*** | (0.01) |
| Hours per week | 0.003* | (0.00) | -0.000 | (0.01) | 0.003 | (0.00) |
| Innovative activities | | | | | | |
|   New product | 0.064 | (0.08) | -0.556** | (0.27) | 0.035 | (0.10) |
|   New ideas | 0.094 | (0.08) | 0.666** | (0.28) | 0.146 | (0.09) |
|   R&D spending in last 3 years | 0.389*** | (0.08) | 0.065 | (0.30) | 0.362*** | (0.10) |
| Sources of financing | | | | | | |
|   Credit line | 0.484*** | (0.08) | 0.070 | (0.28) | 0.475*** | (0.10) |
|   Personal loans outstanding | -0.044 | (0.15) | 0.865* | (0.50) | -0.005 | (0.20) |
| Is competition an obstacle? | -0.028 | (0.04) | -0.006 | (0.13) | 0.014 | (0.05) |
| Marketing and sales orientation | | | | | | |
|   Franchise | 0.651*** | (0.11) | 0.178 | (0.32) | 0.793*** | (0.14) |
|   Marketing intensity (reference category = never) | | | | | | |
|     Rarely (once in a while) | 0.186 | (0.13) | 0.987** | (0.47) | 0.111 | (0.17) |
|     Sometimes (few times per month) | 0.251* | (0.13) | 0.387 | (0.38) | 0.150 | (0.17) |
|     Frequently (few times per week) | 0.340*** | (0.13) | 0.467 | (0.40) | 0.259 | (0.16) |
|     All the time (daily) | 0.379*** | (0.13) | -0.010 | (0.41) | 0.449*** | (0.17) |
| Number of observations | 1400 | | 90 | | 913 | |
| Wald chi$^2$ of model fit | 1281.770*** | | 383.964*** | | 853.306*** | |
| LR test vs single-level model | 67.802*** | | 0.000*** | | 32.638*** | |

*Note* – Dependent variable is log of sales. Standard errors in parentheses. Models estimated using hierarchical linear modeling. Industry fixed effects included in all models. * p<0.10, ** p<0.05, *** p<0.01

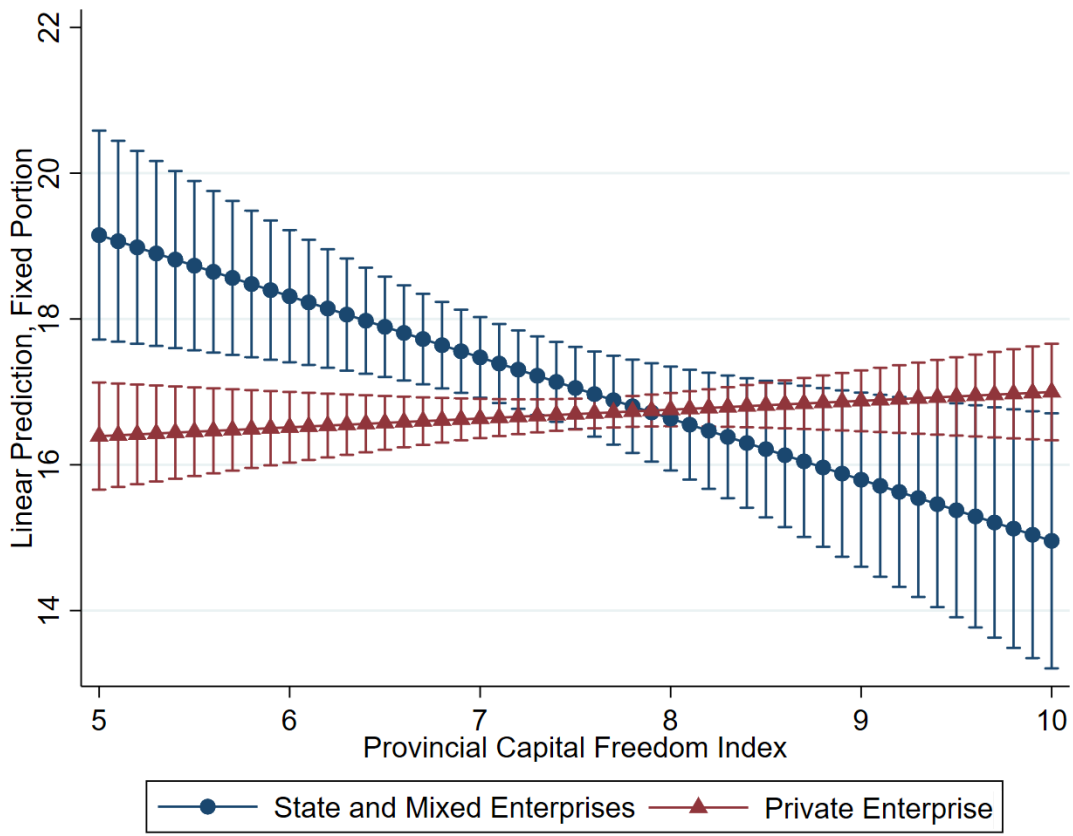

**Fig. 1.** Interaction between privatization and institutional quality on entrepreneurs' sales for young firms (< 5 years of age).